\documentclass[twocolumn,showpacs,preprintnumbers,amsmath,amssymb]{revtex4}

\usepackage{graphicx}
\usepackage{dcolumn}
\usepackage{bm}

\begin{document}


\title{
Effects of polarized organosilane self-assembled monolayers
on organic single-crystal field-effect transistors
}

\author{J. Takeya$^{1,*}$}
\email{takeya@criepi.denken.or.jp}


\author{T. Nishikawa$^{2,3,5}$}
\author{T. Takenobu$^{2,4}$}
\author{S. Kobayashi$^{2,4}$}
\author{C. Goldmann$^6$}
\author{C. Krellner$^6$}
\author{T. Shimoda$^{3,5}$}
\author{T. Mitani$^{5}$}
\author{Y. Iwasa$^{2,4}$}
\author{B. Batlogg$^6$}

\affiliation{
$^1$Materials Science Research Laboratory, CRIEPI, Tokyo 201-8511, Japan}
\affiliation{
$^2$Institute for Material Reasearch, Tohoku University, Sendai 980-8577, Japan}
\affiliation{
$^3$Technology Platform Research Center, SEIKO EPSON Corporation, Fujimi,
Nagano 399-0293, Japan}
\affiliation{
$^4$CREST, Japan Science and Technology Corporation, Kawaguchi 333-0012, Japan}
\affiliation{
$^5$Japan Advanced Institute of Science and Technology, Tatsunokuchi,
Ishikawa 923-1292, Japan}
\affiliation{
$^6$Laboratory for Solid State Physics ETH, CH-8093
Z$\ddot{u}$rich, Switzerland
}%

\date{\today}

\begin{abstract}
The surface conductivity is measured by a four-probe technique
for pentacene and rubrene single-crystals laminated on
polarized and nearly unpolarized molecular monolayers
with application of perpendicular electric fields.
The polarization of the
self-assembled monolayers (SAMs)
shifts the threshold gate voltage,
while maintaining a very low subthreshold swing
of the single-crystal devices (0.11 V/decade).
The results, excluding influences of parasitic contacts
and grain boundaries,
demonstrate SAM-induced nanoscale charge injection up to
$\sim 10 ^ {12}$~cm$^{-2}$ at the surface
of the organic single crystals.
\end{abstract}

\maketitle

Significant efforts are being made to bring
organic field-effect transistors (OFETs)
into practical use,
taking advantages of their potential of low-cost production,
mechanical flexibility
and optical as well as chemical sensitivities \cite{application}.
To promote the development,
some basic questions are to be further elucidated,
such as the intrinsic nature of the transport of charge
induced at the surface of the organic semiconductors.
Following this direction, a growing number of experiments has recently been
reported on the single-crystal devices that avoid
complication due to grain boundaries
\cite{podzorov,takeya,boer,podzorov2,hasegawa,sundar}.
In an early study of the single-crystal OFETs,
Takeya {\it et al.} proposed a method of laminating organic crystals
on SiO$_2$/doped Si substrates by natural electrostatic force \cite{takeya}.
Four-probe measurements on the single-crystal devices
revealed genuine transfer characteristics of the OFETs,
free from parasitic contact effects.
In addition, the laminated single-crystal devices are also useful to
examine whether a new finding in polycrystalline thin-film OFETs
is intrinsic to the semiconductor-SiO$_2$ interface
or is an artifact due to grain-boundaries,
because the process of single-crystal device fabrication
is identical to the commonly studied
bottom-electrode thin-film OFETs.
Note that inter-grain charge dynamics dominates
the field-effect properties
in some thin-film devices \cite{horowitzgrain}

Self-assembled monolayers (SAMs) of neutral organosilane molecules are
often incorporated in thin-film OFETs to
passivate the SiO$_2$ surface before
evaporating the organic semiconductor \cite{gundlach1}.
Very recently, Kobayashi {\it et al.} reported
that the threshold gate voltage
$V_{th}$
is shifted in their polycrystalline thin-film devices
when the neutral SAMs are replaced with
highly electron-affine ones \cite{kobayashi}.
The mechanism of the threshold shift $\Delta V_{th}$,
however, was not clear;
rather large standard deviation among the samples
causes difficulty in quantitative comparison of
$\Delta V_{th}$ among different thin-film materials and
to the calculated dipole moment of the SAMs.
Since the inclusion of grain boundaries and contact resistances
may introduce complications \cite{pernstich}, single-crystal samples are
more appropriate to further study the microscopic mechanism of
the observed shift.
In the present work, we laminated pentacene and rubrene crystals
on both polarized and on nearly unpolarized SAMs, to measure the
field-effect conductivity using the four-terminal method.
We reproduce the clear difference in $V_{th}$ between the devices
with the two different types of the SAMs, which demonstrates that charge is
indeed induced at the surface of the crystal
by the adjacent polarized organosilane molecules.
The amount of the doped surface charge, however,
significantly differs between the two organic materials,
which cannot be explained by simple electronic response to
the polarized substrates.

\begin{figure}
\vspace{-1.5cm}
\includegraphics[width=8cm]{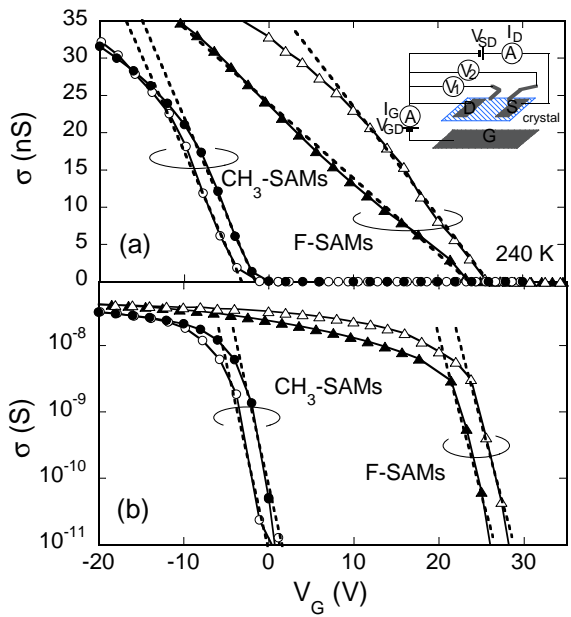}
\vspace{-2cm}
\caption{\label{fig:epsart}
Main panel: Four-probe conductivity vs. gate voltage
of pentacene single-crystal FETs with CH$_3$-SAMs and F-SAMs at 240 K;
(a) a linear plot and (b) a logarithmic plot.
Inset: The circuit diagram of the four-terminal measurement.
The dashed lines are drawn to estimate $\mu$ in (a) and
$S$ in (b).
}
\end{figure}

\begin{figure}
\vspace{-3.5cm}
\includegraphics[width=8cm]{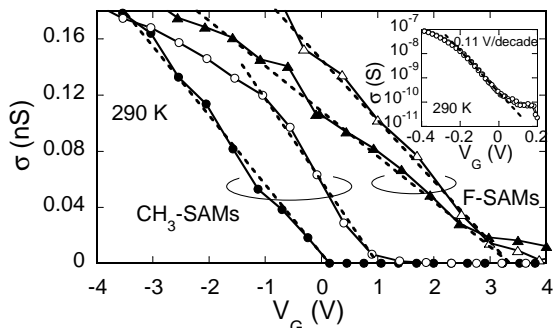}
\vspace{-4cm}
\caption{\label{fig:epsart2}
Four-probe conductivity vs. gate voltage
for rubrene single-crystal FETs.
Main panel: comparison between
CH$_3$-SAMs samples and F-SAMs samples at 290 K. The thickness of the SiO$_2$
is $\sim$ 500 nm.
The inset shows results for a CH$_3$-SAM sample with a SiO$_2$ layer of
less than 100 nm thickness.
The dashed lines were used to estimate $\mu$ in the main panel
and $S$ in the inset.
}
\end{figure}

\begin{table}
\caption{\label{tab:table2}Device parameters of the measured single-crystal FETs.
$d_i$, $\mu$, $V_{th}$, $S$, $D_{it}$ represent thickness of SiO$_2$ dielectrics,
mobility, threshold voltage, subthreshold swing, and interface-trap density,
respectively.}
\begin{ruledtabular}
\begin{tabular}{ccccccc}
Material & SAMs & $d_i$
 & $\mu$ & $V_{th}$ & \hspace{-3mm}
$S$    & $D_{it}$ (10$^{11}$  \\
 &  & ($\mu$m)
 &  (cm$^2$/Vs) &  (V)& \hspace{-5mm} (V/decade)
 &  /cm$^{2}$eV)\\
\hline
pentacene & CH$_3$- & $\sim 0.5$ & $\sim 0.4$ & $\sim -3.5$ & $\sim 1.5$ & $\sim 9$ \\
pentacene & CH$_3$- & $\sim 0.5$ & $\sim 0.4$ & $\sim -1.5$ & $\sim 1.5$ & $\sim 9$ \\
pentacene & F- & $\sim 0.5$ & $\sim 0.15$ & $\sim 24$ & $\sim 2$ & $\sim 20$ \\
pentacene & F- & $\sim 0.5$ & $\sim 0.2$ & $\sim 26$ & $\sim 2$ & $\sim 20$ \\
rubrene & CH$_3$- & $\sim 0.5$ & $\sim 7$ & $\sim 0$ & $\sim 0.4$ & $\sim 2$ \\
rubrene & CH$_3$- & $\sim 0.5$ & $\sim 8$ & $\sim 0.8$ & $\sim 0.7$ & $\sim 4$ \\
rubrene & F- & $\sim 0.5$ & $\sim 4$ & $\sim 3.5$ & $\sim 0.9$ & $\sim 5$ \\
rubrene & F- & $\sim 0.5$ & $\sim 5$ & $\sim 3.5$ & $\sim 1.3$ & $\sim 7$ \\
rubrene & CH$_3$- & $< 0.1$ & $\sim 6$ & $\sim -0.1$ & $\sim 0.11$ & $\sim 2$ \\
\end{tabular}
\end{ruledtabular}
\end{table}

The single-crystal devices are prepared similarly to the processes
described in \onlinecite{takeya}.
In an additional step, before placing the crystals onto the
substrates, we carefully deposited
decyltriethoxysilane or perfluorotriethoxysilane
onto the SiO$_2$ gate dielectric
by chemical vapor deposition (CVD) technique at 150$^\circ$C to form
neutral CH$_3$-terminated SAMs (CH$_3$-SAMs) or electron-affine
F-terminated SAMs (F-SAMs), respectively.
Typical channel dimensions are a length of
around 7~$\mu$m and a width of 40~$\mu$m.
We used the circuitry established in Ref.~\onlinecite{takeya} for the
four-terminal measurements employing
four source-measure units (SMUs) of an Agilent Technology E5207
semiconductor parameter analyzer [Inset Fig. 1(a)].
Since we are interested in the on-off switching,
particular attention is paid to reliable determination of $V_{th}$.
We set $V_D$ to be small enough so that the $I-V$ characteristics
of the channel remain ohmic; otherwise, $V_{th}$ would move to
the positive-$V_G$ direction in the non-ohmic regime
(when the carriers are holes).

In the main panel of Fig. 1(a), we plot
the transfer characteristics
of two CH$_3$-SAM samples and two F-SAM samples.
The horizontal axis represents the gate voltage
$V_G$ applied to the central part in the channel, which is given by
$V_G = V_{GS} + (V_1 + V_2) / 2$.
Hysteresis is negligible at 240 K
when the gate voltage is swept back and forth in helium atmosphere.
Though pronounced hysteresis appears at room temperature,
introducing some ambiguity in the determination of $V_{th}$,
the hysteresis quickly diminishes with decreasing temperature.
The threshold voltage, however, remained essentially
unchanged between 240 K and 290 K.

Fundamental parameters of the samples
are listed in Table 1.
Obviously $V_{th}$ differs
among the two groups
of pentacene single-crystal FETs;
the two CH$_3$-SAM samples switch on at around
0 V, while the two F-SAM samples do around 25 V.
The corresponding difference in the gate threshold field $E_{th}$ is
estimated to be $\sim$ 0.5 MV/cm, which is comparable to
the value $\sim$ 0.7-1 MV/cm reported for
polycrystalline thin-film pentacene FETs \cite{kobayashi,pernstich}.
Noting that the present four-terminal measurements
on the single crystals are free from complications such as
grain boundaries and parasitic contact effects,
the threshold voltage shift is an intrinsic effect
at the interface between the well-ordered SAM molecules
and the almost perfectly ordered pentacene molecules.

The interface quality manifests itself
in sharp on-off switching
as illustrated in Fig. 1(b) on a logarithmic scale.
The subthreshold swing $S$ of
the pentacene devices are around 2 V/decade.
This value indicates that the interface trap density
is at least comparable to or better than that of the best
pentacene thin-film FETs as compared with
literatures \cite{gundlach2,knipp},
taking into account the rather thick dielectric layer in the present
devices ($S C_i$ is evaluated $\sim$ 14 nVF/decade cm$^2$ as renormalized by
the capacitance $C_i$ of the gate dielectrics per area).

In order to examine the universality of the
above observation,
we have also measured the four-probe surface conductivity
of rubrene single-crystal FETs, prepared
in exactly the same way as the pentacene devices.
Figure 2 shows the transfer characteristics of the
rubrene devices with CH$_3$-SAMs or F-SAMs,
which were deposited under the same conditions as for the
pentacene devices.
The mobilities estimated from the slopes in the on-state
are listed in Table 1
together with other device parameters.
The values are comparable to former reports
on both similarly prepared rubrene single-crystal field-effect
devices and devices with softly deposited
polymeric gate insulators \cite{goldmann,podzorov,podzorov2}.
Because of negligible hysteresis of our rubrene FETs,
the comparison of $V_{th}$ is only made at room temperature
this time.

The threshold voltage of the rubrene devices shift
with the intrinsic polarization of the SAMs
as it did for the pentacene samples;
however, the difference in $V_{th}$ is only around 3.5 V, which
is much smaller than $\sim$~25~V for the pentacene devices.
Since we have excluded artifacts from
grain boundaries and contacts to the electrodes,
it is natural to consider that
the electron affinity of the F-SAMs
indeed induces charge redistribution at the
surface of the pentacene and rubrene crystals.
To remove the excess holes, the gate electric field
$E_{th} = V_{th} / d_i$, where $d_i$ is thickness of the
gate insulator, is needed;
in other words, the density of excess holes induced by the
F-SAMs is $p_0 = \epsilon \epsilon_0 E_{th}$,
where $\epsilon \epsilon_0$ is the dielectric permittivity
of the gate insulator.
Taking the values of $V_{th}$ for the pentacene
and rubrene devices,
the excess hole density is estimated to be
$p_0$ $\sim$ $1.1 \times 10^{12}$~cm$^{-2}$
and $\sim$ $1.5 \times 10^{11}$~cm$^{-2}$, respectively,
assuming $\epsilon \sim 3.9$
for the thermally oxidized SiO$_2$.


Sugimura {\it et al.} reported
that surface potential differs by
$\sim$ 0.2 V between the F-SAMs and the CH$_3$-SAMs
deposited on SiO$_2$ based on their Kelvin-probe
measurement \cite{sugimura}.
On the other hand,
the dipole moment of free-standing F-SAMs $\mu$, which is
calculated to be $\sim$ 2 Debye,
would give the potential difference
$\Delta V_s$ of
0.4-1.1 V (depending on assumed SAM density $N$)
as $\Delta V_s = N \mu / \epsilon_{SAM} \epsilon_0$
\cite{kobayashi}.
Note that the latter can overestimate $\Delta V_s$
because it neglects
any depolarization effects as the molecules form the layer.
Taking the smallest value $\sim$ 0.2 V of $\Delta V_s$, still
a gate potential as high as $\sim$ 100 V across our gate dielectric
is necessary to compensate the surface potential induced by the
1-nm thick F-SAMs.
As compared to our experimental results,
$\Delta V_{th}$ is significantly smaller than the above
estimation, indicating presence of
other depolarization mechanism(s)
involving the organic crystals.
Moreover,
the observation that $p_0$ significantly differs between the
pentacene and rubrene devices supports the idea of
the charge rearrangement {\it in} the crystals.
Microscopically, ionic relaxation in response to
the local electric field \cite{northrup,rep} may be
a possible cause of the redistribution of the
(acceptor-type) interface-trap states \cite{volkel,lang}.
Further theoretical and experimental
studies are necessary to fully
elucidate this mechanism of SAM-induced surface charge doping.

Finally, the subthreshold swing $S$ of
the rubrene devices shall be discussed.
For the four devices shown in Fig. 2,
$S$ is evaluated to be
0.4-1.3 V/decade, which is smaller than the values for
the pentacene devices (listed in Table 1).
We have also measured another
rubrene device attached on thinner ($<$ 100 nm) SiO$_2$,
coated with a CH$_3$-SAM.
The inset to Fig. 2 shows the transfer characteristics of this
device.
The slope of 0.11 V/decade measured in this device is the
smallest value reported so far for organic FETs
and is comparable to the value of the best inorganic devices
0.07 V/decade.
This is the result of the high-quality
interface of the rubrene-single crystal devices.
A well-known simple equation relates
$S$ to
density of shallow trap levels (within the thermal energy),
$S = k_B T / e ln(10) (1 + C_{it}/C_i)$,
where $C_{it}$ is the capacitance due to the interface traps
[$k_B$ is the Boltzmann constant and $e$ is the electric charge].
The interface-trap density $D_{it}$ is simply given by $C_{it}/e$ \cite{pope}.
Because of thermal diffusion, the steepness is limited by the
first term in parentheses of the right-hand side of the equation.
As we list in Table 1, $D_{it}$
for the rubrene devices is roughly one order smaller than that of
the pentacene devices.
This estimation, derived from a four-probe measurement,
gives an intrinsic shallow-trap density excluding
artifacts from the contacts.
The difference in $S$ between the pentacene
and rubrene devices indicates that
the interface traps are located inside
the organic materials rather than in SiO$_2$.
For pentacene,
the estimation is consistent with an independent
evaluation by optical measurement \cite{lang}.

In summary,
the incorporation of self-assembled monolayers provides
a variety of new possibilities for
FETs of laminated single crystals.
The highly ordered interface between
the self-aligned silane molecules and
the surface of the molecular crystal facilitates an
outstanding field-effect response,
demonstrated for a rubrene device
as the steepest subthreshold
swing 0.11 V/decade ever reported.
The use of polarized F-SAMs demonstrated that holes are
accumulated
as a result of nanometer scale
charge rearrangement near the interface between the crystal and the gate
oxide.
From a technical aspect,
the present results suggest that
the switching gate voltages can be tuned
by choosing the appropriate SAM molecule from a large number of
available materials.
Combined with
the excellent subthreshold characteristics,
this technique may prove useful for implementation in low-power applications
such as logic components.


The authors acknowledge helpful discussions with D. J. Gundlach
and technical support by S. Mori.
This study was partially supported by a Grant-in-Aid for
Scientific Research (No. 16740214) from the Ministry of Education,
Culture, Sports, Science, and Technology, Japan.
We also thank the Swiss National Science Foundation for financial
support.


\bibliography{penfe}







\end{document}